\documentstyle[aasms4]{article}

\def\keywords{}
\def\acknowledgements{}
\def\apj{ApJ}
\def\baas{BAAS}
\def\mnras{MNRAS}
\def\pasp{PASP}
\def\aap{A\&A}
\def\apjl{ApJ}
\def\apjs{ApJS}
\def\nat{Nature}
\slugcomment{submitted to ApJ Letters}
\begin{document}

\def\macho{{\sc macho}}
\def\newpage{\vfill\eject}
\def\vs{\vskip 0.2truein}
\def\gnu{\Gamma_\nu}
\def\fnu {{\cal F_\nu}}
\def\mass{m}
\def\lum{{\cal L}}
\def\imf{\xi(\mass)}
\def\ilf{\psi(M)}
\def\msun{M_\odot}
\def\zsun{Z_\odot}
\def\met{[M/H]}
\def\vi{(V-I)}
\def\mtot{M_{\rm tot}}
\def\mhalo{M_{\rm halo}}
\def\pp{\parshape 2 0.0truecm 16.25truecm 2truecm 14.25truecm}
\def\la{\mathrel{\mathpalette\fun <}}
\def\ga{\mathrel{\mathpalette\fun >}}
\def\fun#1#2{\lower3.6pt\vbox{\baselineskip0pt\lineskip.9pt
  \ialign{$\mathsurround=0pt#1\hfil##\hfil$\crcr#2\crcr\sim\crcr}}}
\def\ie{{\it i.e., }}
\def\eg{{\it e.g., }}
\def\etal{{\it et al. }}
\def\etalc{{\it et al., }}
\def\kpc{{\rm kpc}}
 \def\Mpc{{\rm Mpc}}
\def\mh{\mass_{\rm H}}
\def\mmax{\mass_{\rm u}}
\def\ml{\mass_{\rm l}}
\def\bc{f_{\rm cmpct}}
\def\br{f_{\rm rd}}
\def\kmsec{{\rm km/sec}}
\def\ibl{{\cal I}(b,l)}
\def\dmax{d_{\rm max}}
%
%--------------------------------------------------------------------
%
%---------------------------------------------------------------------
%

%%%%%%%%%%%%%%%%%%%%
%%%%% put in these lines for electronic submission
\title{The Mass-Function of Low Mass Halo Stars:  
Limits on Baryonic Halo Dark Matter}
\author{David S. Graff\altaffilmark{1} and Katherine Freese\altaffilmark{2}}
\affil{University of Michigan, The Harrison M. Randall Laboratory of Physics,
    Ann Arbor, MI 48109-1120 USA}
\altaffiltext{1}{\tt graff@umich.edu}
\altaffiltext{2}{\tt freese@mich1.physics.lsa.umich.edu}
\received{9 February, 1996}
%%%%%%%%%%%%%%%%%%%%%

%\twocolumn
%\centerline{\bf The Mass-Function of Low Mass Halo Stars:}
%\centerline{\bf Limits on Baryonic Halo Dark Matter}

%%%%%%%%%%%%%%%%%%%%%%
%%%%%%% put in these lines for standard latex
%\bigskip
%\renewcommand{\thefootnote}{\fnsymbol{footnote}}
%\centerline{{\bf David S.~Graff} and {\bf Katherine~Freese}\footnote{\tt graff@umich.edu; freese@mich1.physics.lsa.umich.edu}}
%\vskip 0.1in
%{\it\centerline{University of Michigan, Department of Physics}
%\centerline{Ann Arbor, MI 48109-1120}}
%\centerline{submitted to ApJ Letters 9 February, 1996.}
%\centerline{version of \today}
%%%%%%%%%%%%%%%%%%%%%%%%%%%%%%%%

\begin{abstract}
We derive mass functions (MF) for halo red dwarfs (the faintest
hydrogen burning stars) and then extrapolate to place limits on the
total mass of halo brown dwarfs (stars not quite massive enough
to burn hydrogen).  The mass
functions are obtained from the luminosity function of a sample of 114
local halo stars in the USNO parallax survey (Dahn \etal 1995).  We use
stellar models of Alexander \etal (1996) and make varying
assumptions about metallicity and about possible unresolved 
binaries in the sample.  We find that the MF
for halo red dwarfs cannot rise more quickly than $1/m^2$ as
one approaches the hydrogen burning limit.  Using recent results
from star formation theory, we extrapolate the MF
into the brown-dwarf regime.  We see that likely extrapolations 
imply that the total mass of 
brown dwarfs in the halo is less than $\sim
3\%$ of the local mass density of the halo ($\sim 0.3\%$ for the more
realistic models we consider).  
Our limits apply to brown dwarfs in the halo that come from the same stellar
population as the red dwarfs.
\end{abstract}


\section{\bf Introduction}
\setcounter{footnote}{0}
\renewcommand{\thefootnote}{\arabic{footnote}}

In this paper we determine a mass function for low mass stars
and then extrapolate to derive an upper limit on the number of brown dwarfs.
A key quantity for describing a stellar population is the mass
function $\xi(\mass)$, the density of stars with mass between $\mass$
and $\mass+d\mass$; for $\xi(\mass)$ we use units
\#stars/pc$^3/\msun$.  The mass function (hereafter, MF) 
of halo stars can address the
question of what makes up the dark matter of our Galaxy.
Although it has been
clear for some time that main sequence stars heavier than our sun are
not sufficiently abundant to explain the mass of the Galaxy,
the question has remained whether or not there can
be large numbers of lower mass objects.  Very low mass stars or
substellar objects have been considered the most plausible candidates
for baryonic dark matter.  [White dwarfs, another possibility for
baryonic dark matter, are not considered in this paper.]  Recent work
has ruled out red dwarfs, stars just barely massive to burn hydrogen
($m > 0.092\msun$), as a significant source of dark matter (Graff
\& Freese 1996; Boeshaar, Tyson \& Bernstein 1994; Bahcall \etal
1994).  Brown dwarfs, star-like objects that are just barely too small
to burn hydrogen, are the remaining low mass candidates.  These are
extremely difficult to observe optically (see \eg Burrows and Liebert
1993 for a review) although there have been searches for them
with gravitational microlensing (Alcock \etal 1995, Ansari \etal
1996).  

As our main result, we determine a MF
of halo red dwarfs, and then extrapolate into the brown dwarf regime
 to obtain an upper limit on the
number of halo brown dwarfs.  If brown dwarfs are a primary component of
halo dark matter, then we would expect a steeply rising
MF as we go to decreasing mass.  Instead, we show that
likely extrapolations imply that brown dwarfs are not nearly numerous
enough to be a significant component of halo dark matter.  As a
caveat, let us point out that our limits
apply only to brown dwarfs in the halo that come from the same
stellar population as the red dwarfs (see the 
discussion at the end of the paper).  We also rely
on recent results from star formation. The
goal of our paper is to show that brown dwarfs coming from known
stellar populations are not present in great abundance.
%
%Dark Matter in the Galaxy is primarily in the halo
\footnote{There 
have been several attempts to determine the mass
function of local stars (Miller \& Scalo 1979, 
Kroupa, Tout \& Gilmore 1993, Henry 1995).
Local stars are predominantly Population I disk stars.
A population
II MF for low mass stars from globular cluster NGC 1261 
can be found in Zoccali ${\it et \, al}$ (1995).
In our paper we focus instead on local field halo
stars, which are not in globular clusters.}
%The MF in the halo could in principle be quite different
%from that in globular clusters, in that there is far more dark
%matter in the halo.
%Halo stars presumably formed at the same 
%time as the galaxy.  Since the stars in
%question are old and have low mass, they lie comfortably on the main
%sequence; the precise age of the stars is not an important
%parameter for models of these stars.

Unfortunately, the MF cannot easily be obtained by
observations, since it is difficult to measure the
mass of a star.  Instead, we must convert from the measured quantity,
the luminosity function $\psi(M)$, the number of stars in a magnitude
range $M \rightarrow M+dM$ (note that $M$ refers to magnitude while
$\mass$ refers to mass).  For the luminosity function (hereafter LF), we use
units \#stars/pc$^3/M$.  
%Here we are interested in stars near the
%hydrogen burning limit, where the mass-luminosity relation becomes
%extremely non-linear.  Just above the hydrogen burning
%limit, a red dwarf with a mass of $0.1 \msun$ has a luminosity of
%$\sim 10^{-3}L_\odot$.  Below the hydrogen burning limit, a mature brown
%dwarf with a mass of 0.06$\msun$ has a luminosity three orders of
%magnitude lower, $\sim 10^{-6}L_\odot$ (Burrows, Hubbard, Saumon \& Lunine 1993).
%

%cut
%Even measuring an LF is not trivial.  Ideally, one
%would like to know the distance and absolute magnitude of observed
%stars.  For most data sets of high proper motion stars, observers can
%only measure the relative magnitude and the color of a sample which is
%also contaminated by disk stars.  One approach is
%to construct a series of model stellar populations and to search for
%the best fit to observations (\cite{bs}; \cite{ktg91}; \cite{ktg93}).
%In this paper we instead follow a second approach:
%we use data from trigonometric parallax measurements, which provide
%the distances and absolute magnitudes of stars; from these
%quantities an LF can be constructed.  Parallax measurements give
%the velocities of stars so that one can separate true disk stars from
%halo stars that are temporarily in the disk.  
%Note that we can only
%discuss the local population of halo stars.

To obtain a mass function for red dwarfs, we use the LF from the
extensive parallax survey being carried out by the US Naval
Observatory.  Preliminary results were reported in Monet ${\it et \,
al}$ (1992) with a more complete update in Dahn ${\it et \, al}$
(1995).  They published an LF based on 114 stars with
velocities tangential to the line of sight $V_{\rm tan}\ge 220$km/sec.
These high velocity stars should all be halo stars (temporarily in the disk)
since true disk stars have a local
velocity dispersion of only $\sim$30 km/sec. 
 M\'{e}ra, Chabrier \& Schaeffer
(1996) used this LF to derive an MF as discussed further below; 
see their Fig. 1.  In this paper we present a 
more complete derivation of the mass
function for red dwarfs, and a different interpretation.  We take into account
several potentially important effects not considered by M\'{e}ra et al,
such as unresolved binaries in the
sample and the effects of different choices of metallicity.
In particular, many of the `stars' in the sample are likely to be
unresolved binaries, and substantial numbers of binaries 
may drastically alter the results.
The MF derived by M\'{e}ra et al falls into the range of
possibilities bracketed by the parameter range we consider.
However, our interpretation is different.  We emphasize a different
slope of the MF near the hydrogen burning limit.  (By ``slope'', we
mean the logarithmic slope ${\rm d}\log\xi / {\rm d}\log\mass$.)
%Although our interpretation
%is different, our derived MF is consistent with theirs.
%In addition, we 
In addition we use theoretical results from star formation
to extrapolate to lower masses and limit the abundance of halo
brown dwarfs.  Because we are using parallax data, we can only discuss
the local population of halo stars.
%cut 
% Again, the goal of our paper
%is to show that brown dwarfs coming from known stellar
%populations are not present in great abundance.

\section{Deriving the MF of Red Dwarfs}
As described above, we use the LF obtained from
the parallax survey of Dahn et al (1995).
To convert from the luminosity function $\psi(M_V)$ to the mass
function $\xi(\mass)$, we use
\begin {equation}
\label{lummass}
\xi(m)=\left|{\frac{d M_V}{d\mass}}\right| \psi(M_V).
\end {equation}
Clearly we need a mass-luminosity relation.  We use the models
calculated by Alexander ${\it et \, al}$ (1996). 
In Fig. 8 of their paper, one can see that
the mass-luminosity relation is non-linear.  The luminosity of stars
drops fairly slowly with decreasing mass until
an inflection point at $\mass \approx 0.4\msun$.  Then the
mass-luminosity relation turns over, and the luminosity rapidly becomes
dimmer as the mass decreases toward $\mass_H$.
We fit between the calculated points of (Alexander \etal 1996) using cubic
splines.  Since our MF will depend
on certain assumptions, we will always consider the
most `conservative' case, i.e., we generate the steepest
possible MF as one goes to low mass, corresponding
to the largest number of low mass red dwarfs and of brown dwarfs.

We use the model of Alexander \etal with stellar
metallicity $Z = 3 \times 10^{-4}$.  Nearly all stars in the sample
have larger metallicities than this, as can be seen from the
comparison of parallax data points with theory for various
metallicities in Fig. 6 of Alexander \etal (1996).  We use this value of
$Z$ because it gives the most conservative result
(the steepest MF towards low mass).
One can see this effect in
Fig. 1a for two different metallicities and can understand it
as follows. High Z stars
are dimmer than low Z stars of the same mass, while high
Z stars are more massive than low Z stars of the
same luminosity.  
%By underestimating the metallicity, we are
%misinterpreting high metallicity, high mass stars as low metallicity,
%low mass stars.  
For the range of luminosities we are considering, 
if we underestimate Z, stars would be on 
a much steeper part of the $M_V (m)$ curve and many
would have masses in a narrow mass range
close to the hydrogen burning limit $\mass_H$.  In addition,
the factor $|d M_V/d\mass|$ of Eq. (\ref{lummass}) becomes large for
stars near $\mass_H$ (because the mass range for low Z stars
of a given luminosity is so small).  These two effects
combine to cause the MF to steepen towards lower mass 
as we lower the
metallicity.  Thus, using $Z = 3 \times 10^{-4}$ gives the steepest
allowed MF and the largest number of brown dwarfs.

In addition, the reported LF can differ from the true LF
because of potential errors and biases in the parallax data.
%In addition, potential errors and biases in the parallax
%measurements may lead to an overestimate of the rate that the mass
%function grows toward small masses.  The reported
%LF can differ from the true LF.
%One source of error is the Malmquist bias:
% as
%\begin{equation}
%\label{bias}
%\psi^\prime(M)=\int dM^\prime \ \psi(M^\prime)\
%exp\left(-\frac{(M-M^\prime)^2}{2\sigma^2}\right){\cal G}(M, M^\prime, \sigma).
%\end{equation}
%where $\psi$ is the true LF, $\psi^\prime$ is the
%observed luminosity, $\sigma$ is the r.m.s. uncertainty in the
%absolute magnitude of a star due to random errors in the observation
%and ${\cal G}$ is a function which takes into account the additional
%systematic errors.
%
Observers on
average underestimate the luminosity of the stars because there are
more distant stars than nearby stars (\cite{lk73}).  Also, due
to random errors, there is a diffusion-like broadening of the 
LF (\cite{sip89}) so that the observed LF does not
drop as fast as the true one as one goes to dim stars.  Both systematic 
biases (known as Malmquist bias)
will cause us to have two many small dim stars in our sample, so we
will over-estimate the steepness of the MF as one goes
to small mass.  Again, our results are conservative.
%go in the direction
%of giving the largest number of the lowest mass red dwarfs and of brown
%dwarfs.

%First we use this LF to estimate the mass density of
%red dwarfs in the halo.  Previously several authors (Graff \& Freese
%1996; Boeshaar, Tyson \& Bernstein 1994; Bahcall \etal 1994) pointed
%out that there %are few red dwarfs in the Galactic halo.  These
%authors used data from deep optical searches of the halo which found
%only 5 or 6 red dwarfs.  Using the LF of Dahn \etalc
%we can obtain a more accurate estimate of the mass density of halo red
%dwarfs as follows:
%\begin {equation}
%\rho=\int dM_V \psi(M_V) \mass(M_V) ,
%\end{equation}
%where subscript $V$ refers to visible band.  Because this result is based
%on 114 stars instead of 6, it is much more accurate. 
%We use the mass-luminosity relation
%$m(M_V)$ from Alexander ${\it et \, al}$ (1996).
%We find that the local halo density for red dwarfs is $2 \times 10^{-5} \msun /
%$pc$^3$ for masses between ~0.09-0.4 $\msun$.  For a halo mass density
%of $7 \times 10^{-3} \msun / $pc$^3$ (Alcock \etal 1995), red dwarfs
%are only 0.3\% of the mass of the halo.

%cut
The LF derived by Dahn \etal grows to a maximum value
at $M_V=12$ and then drops off again as one goes to lower mass.
[Hereafter, we use the words rising and falling to refer to the
behavior of functions as one goes to {\it smaller} values of mass or
of luminosity.]  
One might naively interpret this to mean that there
cannot be many stars of low mass.  However, this interpretation may
not be correct because of the shape of the mass-luminosity curve
(\cite{ktg90}); 
%the sharply decreasing derivative of the curve enters in
%Eq. (\ref{lummass}).  For example, consider a rising MF: to
%get the LF, one multiplies the increasing mass
%function by the decreasing $|d\mass/dM_V|$.  One would expect a
5LF that reached a maximum and then turned over for all
%but the steepest MFs.  Indeed 
a decreasing luminosity
function does not necessarily imply a decreasing MF.

{\bf A Complication: Binaries:}
A complication arises in the implementation of Eq. (\ref{lummass}):
%which takes us from a LF to a MF: 
some of the stars in the survey may actually be
%we do not know whether or not any of the stars in the survey are actually
unresolved binaries.  If so, the luminosity of the binary is due to
light from both stars.  If we mistakenly interpret the light to be
only from a single star, we may overestimate the mass of the star.  
To deal with this potential problem, we make a variety of
assumptions about unresolved binaries, and plot
the corresponding mass functions in Fig. 1. 
Our models represent extremes which bracket the possible range
of MFs due to unresolved binaries.
We shall see that for all cases, we still
conclude that brown dwarfs are an insignificant component of the mass
of the halo.

%Below $\mass_1$ and $\lum_1$ refer to the mass and luminosity of the
%primary star (the larger of the two stars), while  $\mass_2$ and $\lum_2$
%refer to the mass and luminosity of the secondary.
%
%A substantial, but unknown fraction of disk stars are binary.  The
%fraction of binary halo stars is even less well understood.  The
%binary stars in the sample cannot be resolved, so we must be aware
%that any of the stars under observation could be binary stars.  If an
%observed star is binary, it may be that the luminosity of the primary,
%$\lum_1$, is as much as 0.75mag less than the observed luminosity of
%the system.  This shift in luminosity will cause a corresponding shift
%in the MF.  This shift depends on the binary fraction $f_b$
%and also on the distribution of mass ratios (the ratio of the mass of
%the primary star to the mass of the secondary star).
%
In Model I we make the most common assumption about binary stars:
the masses of
the two stars are independent and yet are drawn from the same distribution;
i.e., there is a single MF which determines the mass of each star.
Note that we do allow the binary companion to have a mass below
the hydrogen burning limit (we believe that there is no physical reason
to cut off the distribution here).  Thus our model (and its conclusions)
is different from
that of other authors (Kroupa, Tout \& Gilmore 1991, 1993; Piskunov and Malkov 1991)
who do not consider brown dwarf companions.
%\footnote {Our model is somewhat different from that of these authors
%because they do not consider a system containing one brown dwarf and
%one Main Sequence star to be ``binary''.  This will cause us to reach
%different conclusions}.  
We here show
that if the mass density in brown dwarfs is very small for the no
binary case, then it is also very small for the case of model I.
Assume that brown dwarfs are far more numerous than true stars. Then
by our assumption that the masses of the two stars are drawn from the
same MF, the secondary in any binary system is nearly
always \footnote{Since the above 
authors did not allow brown dwarf companions, 
this statement would not be true for their work; our model II more
closely approximates their results.} a brown dwarf
\footnote{There has been an intensive search for
brown dwarfs as binary companions.  This search has only found two
brown dwarfs despite examining hundreds of systems; thus brown dwarfs
make up $<1\%$ of the mass of disk stars (Rebolo, Zapatero Osorio \&
Martin 1995; Nakajima \etal 1995).  Because brown dwarfs cool as they
age, one can only hope to optically detect young brown dwarfs.  Thus,
these optical searches are not sensitive to a halo population of brown
dwarfs.}.  The contribution of the brown dwarf to the total luminosity
of the system is negligible and we can derive the mass of the primary
from the luminosity of the system.  This means that the MF
derived using a single star mass-luminosity relation also describes
the MF for binary stars.  But, as we will show below, this single
star MF implies that there are not large numbers of brown
dwarfs, contradicting our assumption.  Thus, if the mass density in
brown dwarfs is very small for the no binary case, then it is also
very small for the case of model I. 

%if the masses of the two stars are chosen
%independently from the same distribution, there cannot be many more
%brown dwarfs than red dwarfs.

Although model I is the most plausible, it is difficult to {\it directly} 
calculate the actual mass function from the observed massfunction (although 
the reverse process is trivial).
Thus we tried two other binary star models.  Model II is
plausible.  Model III is extremely unlikely, and is designed to
exaggerate the shift in slope of the MF.  In both models,
we analytically determined the MF which would produce the
observed LF and which would satisfy the assumed
binary composition.  Both models steepen the MF towards
low masses.  [Note that the direction of this effect is in agreement
with what was found by Kroupa, Tout, \& Gilmore (1991, 1993), who studied the effects
of unresolved binaries in other data sets (typically Pop. I)].

In model II, each ``star'' has a probability $f$ (the binary fraction)
of being an unresolved binary.  In every binary system,
we take the mass of the secondary $m_s$ to be evenly distributed
between zero and the mass of the primary $m_p$; i.e.,
the binary mass ratio $q = {m_s \over m_p}$ 
has an equal chance of being anywhere in the interval [0,1].
For each bin $M_V$ in the observed LF histogram
$\psi\prime(M_V)$, a fraction $(1-f)\psi^\prime(M_V)$ of the LF is
assumed to represent single stars whose absolute magnitude has been
accurately determined.  The remaining fraction $f\psi^\prime(M_V)$
represents all unresolved binaries and is divided into 100 parts $i$,
each having a mass ratio $q_i=i/100$, where $i$ is an integer
between 0 and 100.  For each part $i$ we find the
two masses $m_{p,i}$ and $m_{s,i}=q_i m_{p,i}$ with visual magnitudes
$M_{Vp,i}$ and $M_{Vs,i}$ such that the combined luminosity of the system
is exactly what is observed. We then put these
pieces together to obtain the luminosity function $\psi(M_V)$ corrected 
for unresolved binaries; the results from model II will be discussed shortly.
%\begin{equation}
%\label{modelii}
%\psi(M_V')=(1-f)\psi^\prime(M_V)\ +\ f\sum_{M_V^\prime}\sum_{i=0}^{100}\psi(M_V^\prime)(\delta_{M_V,M_{V1,i}}+\delta_{M_V,M_{V2,i}}.
%\end{equation}
%where the $\delta$'s distribute the old LF into the
%bins of the new LF depending on how close each
%$M_{V,i}$ is to the center of a particular bin $M_V$.

In model III, we assume that all of the stars are in binary systems,
and that the two stars have the same luminosity.  Thus the luminosity
of the primary is half that of the system.  
Compared to the no-binary case, 
in model III the luminosity of each star is half as much,
the magnitude is larger by an amount 0.75, and the mass of each star
is smaller.  Also, there are twice as many stars
because each system is a binary.
%underestimate the magnitude of each star by an amount 0.75,
%overestimate the mass of each star, and underestimate
Thus, according to model III, the ``true''
LF $\psi(M_V)$ can be derived from the
observed LF $\psi^\prime(M_V)$ by:
%\begin{equation}
%\label{modeliii}
$\psi(M_V)=2 \times \psi^\prime(M_V-0.75)$.
%\end{equation}
Although the shape of the LF is unchanged, this model
generates a steeper MF for the same reasons a lower
metallicity model generates a steeper MF:  stars lie
on a steeper part of the $M_V(m)$ curve so that their masses lie in the
narrow mass range near the hydrogen burning limit, and the factor $|d
M_V/d\mass|$ of Eq. (\ref{lummass}) becomes large as discussed in
section 2. 

Our results for all the models are plotted in Fig. 1.  
The MF of red dwarfs has been obtained from the LF using
Eq. (\ref{lummass}) for four different cases. 
%cut (This is redundant with the figure caption)
%In each panel we have
%plotted as our `standard' model the case of $Z=3\times 10^{-4}$ and no
%binaries (indicated by crosses with no error bars).  The first panel
%also plots the MF for $Z = 6 \times 10^{-4}$ for
%comparison: lower $Z$ leads to a steeper MF as described
%previously.  The other two panels use $Z=3\times 10^{-4}$ and binary
%models II and III. For the binary models, we
%plot the total MF (primaries and secondaries).    
%Model II is shown with two different binary
%fractions, $f=\{0.5,1.0\}$.
Both binary models II and III generate
MFs that are steeper than if we ignored unresolved
binaries.  In the plots, the MF has been multiplied by
$m^2$ so that one can most easily see how it approaches the hydrogen
burning limit (as described in the next section).  Our result is that,
at the low mass end,
the MF behaves as $\imf \sim m^{-\alpha}$ with $\alpha
\le 2$ for all four cases we consider.  In fact, only binary model III
could have $\alpha=2$ while all other cases have $\alpha<2$.

For comparison, M\'{e}ra et al (1996) found the MF
only for the case of no binaries, and roughly estimated $\alpha \sim 2$.
They obtained this number by fitting to the MF over a large
mass range.  However, we find that for this case of no binaries,
the slope $\alpha$ is smaller near the hydrogen burning limit, 
as one can see from Fig. 1a; it is this smaller slope at
the lowest mass end of the red dwarf range that is relevant
for extrapolating into the brown dwarf regime\footnote{M\'{e}ra
{\it et al} (1996) found $\alpha \sim 2.5$ for the data
of Richer and Fahlman (1992); however these data had
difficulty distinguishing least luminous stars from galaxies
and disagree with the results of Hubble Space Telescope (Bahcall
${\it et al}$ 1994, Graff and Freese 1996).}.
Below we also take advantage of recent
results from star formation theory to guide our extrapolation, 
and this leads us to a different interpretation
of the importance of the shape of the MF near the hydrogen
burning limit.  Also,
our work has considered a wider range of possibilities for binaries
with the aim of obtaining a careful estimate of the range of
possibilities for $\alpha$.  As seen in the next section, the fact
that we find a slope which {\it always} satisfies $\alpha \leq 2$
has important consequences.

\section{Upper Limit on the Total Mass of Halo Brown Dwarfs}

{\it \bf Extrapolating the MF to Brown Dwarfs --
Theoretical Input:} Given our mass
functions of halo red dwarfs, we will now extrapolate to the lower mass
brown dwarfs.  This extrapolation should be guided by theory.  The
work of Adams and Fatuzzo (1996) predicts the mass of a star based on
the physical properties of the cloud core that forms that star.  In
their theory, stars determine their own masses through strong stellar
winds and outflows.  They predicted two extremes for the MF of stars
forming in molecular clouds.  At one extreme, the stellar mass is
determined by one factor, \eg the sound speed of the molecular cloud
core.  In this extreme, the mass function of stars would be a power law.
%cut
%
%I think that if the reader wants more information, he can look
%in Fred & Marco's paper
%
%  Not all cores have the same value of that factor; instead there
%is a distribution of values.  If that distribution is power law, as
%might be expected in turbulent systems with no natural scale for the
%quantity, then the distribution of stellar masses, i.e., the mass
%function, is also likely to be a power law.  
%On a log-log plot (and in
%Fig. 1), the MF should then be a straight line.
At the other extreme, the mass of the star is determined by 
a large number of independent physical variables.
%cut
%  The 
%mass of the star is proportional to the
%product of various powers of the initial conditions which formed the
%star: $\mass=A \prod_i \alpha_i^{\gamma_i}$ where the $\alpha$'s
%stand for the sound speed, rotation rate, etc., i.e.,  
%all the physical parameters present at the birth of the
%star.  Thus, $\log \mass=\log A + \sum_i \gamma_i \log \alpha_i$.
%Here $\log \mass$ is the sum of randomly distributed terms.
Invoking the central limit theorem, Adams and Fatuzzo reasoned that
$\log \xi(\mass)$ should be a Gaussian; \ie the mass should
be log-normal distributed\footnote{To leading order, the MF 
of disk stars appears to be fit by a log-normal (Miller \& Scalo 1979).}.
%[A log-normal MF would appear as
%a parabola on a plot of $\log \xi(\mass)$ vs. $\log \mass$ and in 
%Fig. 1].

These two cases represent extreme limits.  We expect the actual
function to lie somewhere between the two.  In short, a MF
should {\it not} be concave up, only flat or concave down on a log-log
plot.  There are other current theories of star formation based on
completely different physical reasoning, such as heirarchical
fragmentation (\cite{zinnecker}) and collisions between
cloud cores (\cite{pp95}).  All of these theories predict mass
functions that rise less quickly than power laws, and thus lie within
the same range of possibilities as discussed above.
All the possible MFs generated by our different models are
consistent with a mixture of the two theoretical models over the red
dwarf mass range, \ie all are flat or concave down.
% Since the
%observed MF has the same basic form as the theoretical
%prediction for red dwarfs, we feel justified in using this theory to
%extend the MF to include brown dwarfs.

{\it\bf Upper limit on the total mass in Halo brown dwarfs:}
The total mass in brown dwarfs is
\begin {equation}
\label{total_mass}
m_{\rm tot}=\int_0^{m_H} \mass \xi (\mass) d \mass.
\end{equation}
Since there are so few red dwarfs relative to the halo, the total mass
of brown dwarfs can only be large if the MF is steeply rising
as one goes to small masses.
In the theory discussed above, any possible extrapolation of an
MF lies between two possible extremes: power law or
log-normal.  Within these
two bounds, there will be more brown dwarfs if we extrapolate with
a power law.  Hence we use $\xi \propto \mass ^{-\alpha}$ to place an
upper bound on the total mass of brown dwarfs in the halo.
Note that we extrapolate the slope at the low mass end of the
observed mass function.

If $\alpha<2$ then the integral in equation (\ref{total_mass}) converges
and one finds the total mass in brown dwarfs to be of the same order
of magnitude as the total mass in red dwarfs.  If $\alpha \ge 2$, then
the integral diverges.  Since the mass density of brown
dwarfs is not infinite, we mean by this divergence that the total mass
density is dependent on a lower mass cutoff of the MF.  If
$\alpha=2$ then the integral diverges, but only logarithmically, \ie
each order of magnitude of mass range contains an equal total mass.
Thus, even for a lower limit of $10^{-7}\msun$, 
the total mass in brown dwarfs $\sim 12$ times the mass in
red dwarfs, or only 3\% of the local halo mass.  If $\alpha>2$, the
integral diverges fast enough that there could be enough brown dwarfs
to fill the halo.

As shown in Fig. 1, for all of our models, with varying metallicities
and binary content, a power law extrapolation to low masses has a
power of $\alpha\le 2$. Thus brown dwarfs cannot contribute more than $\sim
3 \%$ of the local halo mass.  
Only our most extreme model (binary model III) can have $\alpha=2$; the more
sensible models have $\alpha<2$.  Eq. (\ref{total_mass}) with
$\alpha<2$ implies that there cannot be many more
brown dwarfs than red dwarfs.  Previously we estimated that red dwarfs
contribute $\sim 0.3 \%$ of the halo (for a halo mass density of $7
\times 10^{-3} \msun/pc^3)$.  In summary, brown dwarfs cannot
contribute more than $\sim 3\%$ of the mass of the halo, and in
realistic models the limit is an order of
magnitude smaller.

\section{Discussion}
Red dwarfs are a tiny fraction, ~0.3\%, of the
mass of the galactic halo (\cite{gf96}).  In order for any appreciable
fraction of the halo mass to consist of low mass star-like
objects, there must be substantially more mass in brown dwarfs than in
red dwarfs.  We estimated the mass in brown dwarfs by extrapolating
the mass function (MF) of red dwarfs below the hydrogen burning limit.  Our
extrapolation was based on current theories of star formation, which
predict that the MF is either flat or concave down on
a log-log plot (or in fig. 1), {\it i.e.}, the MF rises
no faster than a power-law.  We made several assumptions about the
metallicity and binary composition of the stars in the sample.
% All assumptions yielded MFs
%which imply that there are not many more brown dwarfs than red dwarfs.
We found that likely extrapolations imply that brown dwarfs make up less than $\sim3\%$ of the local mass
density of the halo.  For our most realistic models, the limit on the 
total mass in brown dwarfs is roughly the total mass in red dwarfs,
$\sim 0.3\%$ of the local mass density of the halo.

To repeat an earlier caveat: our limits on the brown dwarf density
assume that the brown dwarfs in the halo come from the same stellar
population as the red dwarfs.  The possibility always
remains that there are large numbers of brown dwarfs from an entirely
different population of stars (Population III).  However, theoretical
work on star formation (Adams and Fatuzzo 1996) indicates that an
earlier population of stars is likely to be skewed towards a
predominance of high mass stars, not brown dwarfs.  There may also be
some yet undiscovered mechanism of star formation which allows the
MF to rise faster than power-law.  
In this paper we have presented red dwarf mass functions and have
shown that brown dwarfs which are members of known stellar populations are
not present in the halo in great abundance.

\acknowledgements We thank Santi Cassisi for sharing unpublished work
with us.  We also thank F. Adams, G. Basri, P. Kroupa and J.
Bahcall for useful discussions.  We acknowledge support from 
NSF PHY-9406745 and from the Univ. of Michigan Physics Dept.
\pagebreak

\pagebreak
{\bf F I G U R E  \,  C A P T I O N S}

Figure 1: The mass function of red dwarf halo stars.  Each of the four
models is derived from the LF of Dahn ${\it et \,
al}$ (1995) but assumes different metallicity and binary content.  In
all three panels, crosses without errorbars illustrate the mass
function derived for stars with metallicity Z = $3 \times 10^{-3}$ and
no binary companions.  Error bars are due to Poisson errors in the
LF.  The other model presented in panel (a) has Z =
$6 \times 10^{-3}$ (no binaries) for comparison.  Lower metallicity
leads to a MF with more low mass stars.  Panels (b) and (c)
show binary models II and III for $Z = 3 \times 10^{-3}$ as described
in the text and can be compared with the no binary model (crosses) of
the same metallicity.  For binary models we plot the total MF
(primaries and secondaries).
Panel (b) shows model II for two different
binary fractions, $f=\{0.5,1.0\}$; $f=0.5$ is
closer to the no-binary model (crosses).   
Binary model II has a somewhat
steeper MF than the no binary case.  Binary model III has
been designed to exaggerate the number of low mass stars compared to
high mass ones and is unrealistic.

We have multiplied the vertical axis by $\mass^2$ to emphasize
that all MFs converge at low mass
(see eq (\ref{total_mass}) and subsequent text).
A MF which is decreasing to the left will 
converge; one that is increasing will diverge.  
One that is flat will diverge, but only logarithmically.  Even for 
the extreme case of binary model III, the MF
is flat or decreasing and the total 
mass in brown dwarfs could be at most $\sim 3\%$. For more likely
models I and II, the limit on the total mass in brown dwarfs is 
very roughly the total mass in red dwarfs, 
$\sim 0.3\%$ of the local mass density of the halo.


\begin {thebibliography}{DUM}

\bibitem[Adams \& Fatuzzo 1996]{af96} Adams, F. C. \& Fatuzzo, M. 1996
\apj,  in press.

\bibitem{macho}Alcock, C. \etal 1996, \apj {\bf 461} 84.

\bibitem[Alexander \etal 1996]{models96} Alexander, D. R., Brocato,
E., Cassisi, S., Castelliani, V., Ciaco, F., \& Degl'Innocenti,
S. 1996 submitted to \aap.

\bibitem[Ansari \etal 1996]{eros96} Ansari, R. \etal 1996 submitted to \aap.

\bibitem[Bahcall \etal 1994]{bfgk94} Bahcall, J., Flynn, C., Gould,
A. \& Kirhakos, S. 1994, \apj, {\bf 435}, L51.

%\bibitem[Bahcall \& Soneira 1980]{bs} Bahcall, J. \& Soneira, R. M. 1980
%\apjs, {\bf 44}, 73.

\bibitem[Boeshaar, Tyson \& Bernstein 1994]{btb94} Boeshaar, P. C.,
Tyson, J. A. \& Bernstein, G. M., 1994, \baas, {\bf 185}, \#22.02.

\bibitem[Burrows, Hubbard, Saumon, \& Lunine 1993]{bhsl93} Burrows,
A., Hubbard, W. B., Saumon, D. \& Lunine, J. I., 1993, \apj, {\bf406}, 158.

\bibitem{burrows} Burrows, A. \& Liebert, J. 1993, RMP, {\bf 65}, 301.

\bibitem[Dahn \etal 1995]{usno} Dahn, C. C., Liebert, J, Harris,
H. C. \& Guetter, H. H. 1995, in Proceedings of the ESO workshop ``The
bottom of the Main Sequence and Beyond'' ed. C.G.Tinney
(Springer-Verlag, Heidelberg) 239.

\bibitem[Graff \& Freese 1996] {gf96} Graff, D. S. \& Freese, K. 1996
\apjl, {\bf 456}, L49.

\bibitem[Henry 1995]{henry95} Henry, T. 1995, in Proceedings of the
ESO workshop ``The bottom of the Main Sequence and Beyond''
ed. C.G.Tinney (Springer-Verlag, Heidelberg) 79.

\bibitem[Kroupa, Tout \& Gilmore 1990]{ktg90} Kroupa, P., Tout,
C. A., Gilmore, G. 1990 \mnras, {\bf 244}, 76.

\bibitem[Kroupa, Tout \& Gilmore 1991]{ktg91} Kroupa, P., Tout,
C. A., Gilmore, G. 1991 \mnras, {\bf 251}, 293.

\bibitem[Kroupa, Tout \& Gilmore 1993]{ktg93} Kroupa, P., Tout,
C. A., Gilmore, G. 1993 \mnras, {\bf 262}, 545.

\bibitem[Lutz \& Kelker 1973]{lk73}Lutz, T. E., \& Kelker,
D. H. 1973, \pasp, {\bf 85}, 573.

\bibitem[Miller \& Scalo 1979]{ms}Miller, G. E. \& Scalo, J. M. 1979 \apjs,
{\bf 41}, 513.

\bibitem[M\'{e}ra, Chabrier \& Schaeffer]{mcs96}M\'{e}ra, D.,
Chabrier, G. \& Schaeffer, R. 1996 Europhys. Lett., {\bf 33}, 327.

\bibitem{monet}Monet, D. \etal 1992 AJ, {\bf 103}, 639.

\bibitem[Nakajima \etal 1995]{bd1}Nakajima, T., Oppenheimer, B. R.,
Kulkarni, S. R., Golimowski, D. A., Matthews, K. \& Durance,
S. T. 1995 \nat, {\bf 378}, 463.

\bibitem[Piskunov]{pm91} Piskunov \& Malkov, 1991 Astronomy and
Astrophysics, {\bf 247}, 87.

\bibitem[Price \& Podsiadlowski 1995]{pp95} Price, N. M. \&
Podsialdlowski, P. 1995 \mnras {\bf 273}, 1041.

\bibitem[Rebolo, Zapatero Osorio \& Martin 1995]{bd2} Rebolo, R.,
Zapatero Osorio, M. R. \& Martin, E. L. 1995 \nat, {\bf 377}, 129.

\bibitem[Richer, and Fahlman, 1992]{rf92} Richer, H.B. and Fahlman,
G.G. 1992 Nature, {\bf 358}, 383.

\bibitem[Stobie, Ishida \& Peacock 1989]{sip89} Stobie, R. S., Ishida,
K. \& Peacock, J. A. 1989, \mnras {\bf 238}, 709.
 
\bibitem[Zinnecker 1984]{zinnecker} Zinnecker, H. 1984 \mnras {\bf
210}, 43.

\bibitem[Zoccali \etal 1995]{zoc95} Zoccali, M., Piotto, G., Zaggia,
S. R. \& Capaccioli, M. 1995, in Proceedings of the ESO workshop ``The
bottom of the Main Sequence and Beyond'' ed. C.G.Tinney
(Springer-Verlag, Heidelberg) 261.

\end {thebibliography}
\end {document}
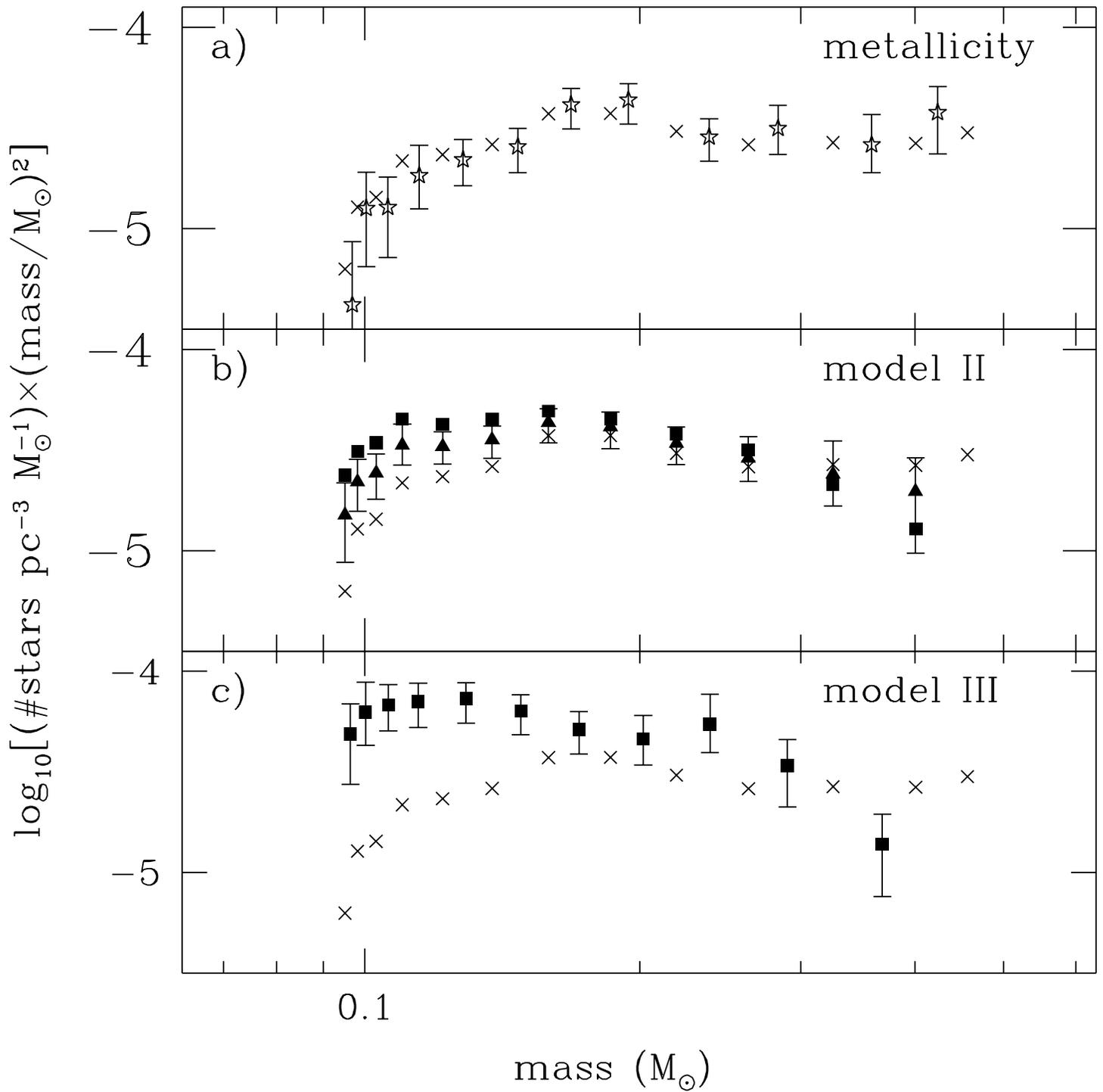
